\documentclass[twoside,twocolumn,english,prl,showpacs]{revtex4-1}
\usepackage{babel}
\usepackage{graphicx}
\usepackage{dcolumn}
\usepackage{amsmath}
\usepackage{stmaryrd}
\usepackage{float}

\begin{document}
 
\title{Nonlinear Fano Profiles in the Optical Second-Harmonic Generation from Silver Nanoparticles.}

\author{J. Butet, G. Bachelier}
\email[Corresponding author: ]{guillaume.bachelier@grenoble.cnrs.fr; Present address: Institut N\'eel, CNRS et Universit\'e Joseph Fourier, BP166, 25 rue des Martyrs, 38042 Grenoble, France.}
\author{I. Russier-Antoine, F. Bertorelle, A. Mosset, N. Lascoux, C. Jonin, E. Benichou and P.-F. Brevet}
\affiliation{Laboratoire de Spectrom\'etrie Ionique et Mol\'eculaire, Universit\'e Claude Bernard Lyon 1 - CNRS (UMR 5579), Bat. A. Kastler, 43 Bd du 11 novembre 1918, 69622 Villeurbanne, France.}

\begin{abstract}

The resonance effects on the optical second harmonic generation from 140 nm silver nanoparticles is studied experimentally by hyper-Rayleigh scattering and numerically by finite element method calculations. We find that the interferences between the broad dipolar and narrow octupolar surface plasmon resonances leads to nonlinear Fano profiles that can be externally controlled by the incident polarization angle. These profiles are responsible for the nonlinear plasmon-induced transparency in the second harmonic generation.

\end{abstract}


\maketitle

The optical response of noble-metal nanoobjects is dominated by the surface plasmon resonance (SPR) associated with the collective motion of conduction electrons \cite{Maier2007}. The latter leads to enhancements of the electric field in the vicinity of the nanostructure with a length scale shorter than that allowed by diffraction limit \cite{Schuller2010}. Recently, a new interest has been paid to SPRs owing to the appearance of Fano interferences \cite{Fano1961,Miroshnichenko2010} arising from the hybridization of a subradiant and superradiant plasmon mode \cite{Hao2008,Lukyanchuk2010,Gallinet2011} or from the interference between a single SPR and the interband transition continuum~\cite{Bachelier2008P,Shegai2011}. The opportunities offered by these unique lineshapes are promising for many applications in sensing, owing to their inherent sensitivity to local changes 
in geometry and environment, or in the plasmon-induced transparency, where the destructive interferences can induce a transparency window \cite{Anker2008,Lukyanchuk2010,Zhang2008,Liu2010}. Still, the ability to macroscopically and reversibly control these Fano profiles is largely unexplored.

A potential route toward this issue might be given by nonlinear optics. From this point of view, the second harmonic generation (SHG), a process where two photons at the fundamental frequency are converted into a single photon at the harmonic frequency, has an ambivalent behavior since, despite the potential plasmonic enhancement, it is forbidden in the dipolar approximation in centrosymmetric materials such as noble metals \cite{Sipe1980,Dadap1999}. Unless noncentrosymmetric shapes are chosen \cite{Kauranen2007,Valev2010,Gentile2011,Halas2011}, this leads to very weak signals, although measurements at the single nanoparticle (NP) level have been achieved already \cite{Schon2010,Butet2010N,Jin2005}. The harmonic photons are generated either on the particle surface, where the centrosymmetry is broken (local response), or in the bulk due to the field gradient (nonlocal response). The recent observation of interferences between the selected dipolar and octupolar modes, in the sense of Mie multipoles, has opened the way to a quantitative determination of the relative efficiencies of these nonlinear sources in spherical plasmonic NPs \cite{Butet2010P,Bachelier2010}.

In this letter, we investigate both experimentally and theoretically the resonance effects on the SHG by 140~nm silver NPs. In contrast with previous works~\cite{Hao2002,Ray2010,Butet2010P,Bachelier2010}, we analyze how the interference contrast between the nonlinearly excited dipolar and octupolar plasmon modes is modulated while varying the incident wavelength. By comparing the experimental results with finite element method (FEM) simulations, nonlinear Fano profiles are revealed as arising from the interaction of the broad tail of the dipolar SPR and the relatively narrow octupolar SPR at the harmonic wavelength. The typical asymmetric line shapes are observed both in the local surface and nonlocal bulk contributions to the SH intensity. Their sensitivity to the incident polarization is addressed evidencing a nonlinear version of the plasmon induced transparency phenomenon \cite{Lukyanchuk2010,Zhang2008,Liu2010}.

We use nearly spherical silver NPs synthesized as in Ref.~\onlinecite{Chen2010}. Their size and shape distribution are narrowed by centrifugation and checked by transmission electronic microscopy. The extinction spectrum of the NP solution, obtained by UV-visible spectroscopy, is shown in Fig.~1. It is well reproduced by the Mie theory, taking into account a mean diameter of 140 nm and a standard deviation of 12 nm. The Palik tables are used for the dielectric function of silver \cite{Palik1998} and a 1.33 refractive index is chosen for water. As shown in Fig.~1, three distinct SPRs show up due to the dipolar \((l=1)\), quadrupolar \((l=2)\) and octupolar \((l=3)\) modes. The spectral shift and broadening of the dipolar and quadrupolar modes are mainly due to retardation effects arising from the rather large particle size compared to visible wavelengths. Hence, the spherical silver particles are good candidates to investigate resonance effects on high-order plasmon modes around 400~nm, corresponding precisely to the harmonic wavelength range accessible in our SHG measurements.

\begin{figure}[tp] 
\center 
\resizebox*{8cm}{!}{\includegraphics{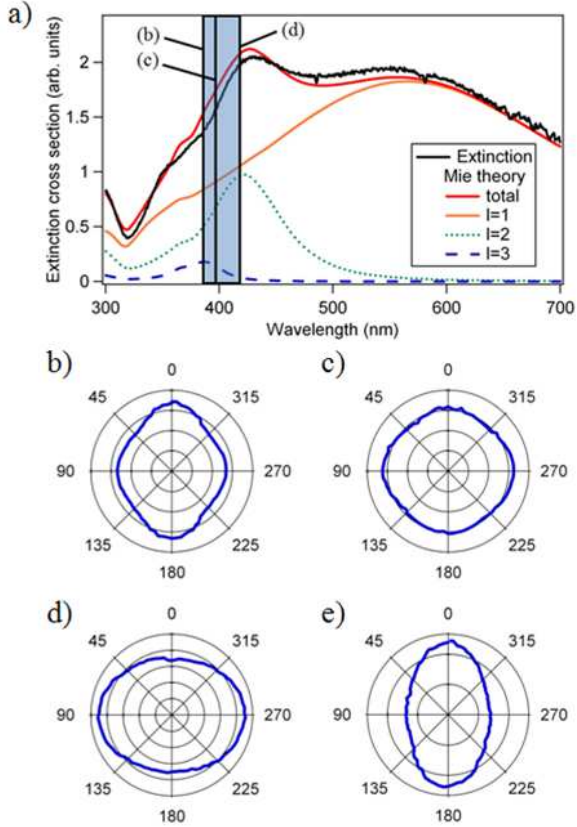}}
\caption{\label{Figure1}(a) Extinction spectrum of the silver NP solution and the corresponding Mie calculation for spherical particles with a mean diameter of 140 nm and a standard deviation of 12 nm. The contribution of the dipolar \((l=1)\), quadrupolar \((l=2)\) and octupolar \((l=3)\) modes are included. The blue area corresponds to the spectral range where SHG measurements are performed. (b-e) Polar plot of the SHG intensity (see text) as a function of the input polarization angle for an harmonic wave polarized in the scattering plane and harmonic wavelengths equal to 383 nm (b), 395 nm (c) and 419 nm (d). The result obtained with 150 nm gold NPs at the harmonic wavelength of 400 nm is shown for comparison (e).}
\end{figure}

The SHG from the silver particles in solution is recorded using a hyper Rayleigh scattering (HRS) set-up. The HRS labeling refers here to the incoherent superposition of the particle responses, owing to their random spatial distribution in the solution, and not to any point-like behavior. The laser source is a mode-locked Ti:sapphire laser delivering pulses of duration around 180 femtoseconds. The polarization angle \(\gamma\) of the input beam is selected with a rotating half-wave plate, \(\gamma=0\;[\pi]\) corresponding to a vertically polarized incident wave. The fundamental beam is focused into a quartz cell with a microscope objective (\(\times 16\), NA~0.32). A red filter is placed in front of it in order to remove any residual light at the harmonic frequency generated prior to the cell. The SH photon collection is performed perpendicularly to the incident beam with a 25~mm focal length lens with a numerical aperture of~0.5. The SH photons are selected in polarization by an analyzer and collected by a cooled photomultiplier after spectral selection by means of a blue filter and a monochromator. A mechanical chopper of the fundamental beam allows rejecting the noise due to a photon counting system. 

The multipolar plasmon contribution to SHG can be quantified either using resonance effects while scanning the excitation wavelength or investigating the radiation patterns by rotating the incident polarization. For this propose, the SH intensity is recorded as function of the incident polarization angle \(\gamma\) for a harmonic wavelength ranging from 383 nm to 419 nm (blue area in Fig.~1a). The corresponding polar plots, associated with SH waves polarized in the scattering plane, are shown for an emission close to the quadrupolar (419 nm, Fig.~1.d) and the octupolar (383 nm, Fig.~1.b) resonances and for an intermediate case (395 nm, Fig.~1.c). Clear deviations from a uniform radiation pattern are observed for emission at both resonant wavelengths. These results can be compared to similar investigations performed recently for gold NPs, i.e. not in resonant conditions (see Fig.~1e and Ref.~\onlinecite{Butet2010P}). They were interpreted in terms of interferences between the dipolar and octupolar modes; the intensity maxima and minima corresponding to constructive and destructive interferences, respectively. 

From the Mie theory, one can describe the SH scattered wave using two kinds of transverse modes: the ``electric'' \(\mathbf{\nabla}\times\mathbf{\nabla}[z_l(kr)Y_l^m(\theta,\phi)\mathbf{r}]\) and the ``magnetic'' \(\mathbf{\nabla}[z_l(kr)Y_l^m(\theta,\phi)\mathbf{r}]\) modes where \(z_l(kr)\) are spherical Bessel functions and \(Y_l^m(\theta,\phi)\) spherical harmonics with \(-l\leq m \leq l\). In non-magnetic media, the SPRs are assigned to the electric modes only, since the magnetic ones do not have an electric field component normal to the particle surface and therefore no surface charge. It is then easy to show that the odd SPRs contribute to the SH wave polarized into the scattering plane, for a collection at right angle with respect to the incident beam, whereas even SPRs can be observed only for a polarization perpendicular to the scattering plane. As a consequence, the quadrupolar (\(l=2\)) resonance plays no role in the measurements reported in Figs.~1b-1d. Indeed, the evolution of the polarization response can be, in a first approach, interpreted by considering that the phase of the electric field associated to the octupolar mode changes sign while scanning the corresponding resonance: constructive and destructive interferences are then exchanged (see Figs.~1b and 1d). In addition, similar responses are obtained for gold and silver particles for excitation above their respective octupolar resonances (compare Fig.~1b and 1e although the emission in Fig.~1b is not yet above the octupolar resonance).

In order to gain more insight into the physical origin of the SHG evolution with wavelength, we have performed FEM calculations following the numerical protocol described in Ref.~\onlinecite{Bachelier2008O}. Here both local surface and non-local bulk contributions are included using the following expressions for the nonlinear polarizations \cite{Bachelier2010}: 
\begin{eqnarray}
P_{surf,\perp}(\mathbf{r},2\omega)&=&\chi_{\perp\perp\perp}E_{\perp}(\mathbf{r},\omega)E_{\perp}(\mathbf{r},\omega),\\
P_{surf,\parallel}(\mathbf{r},2\omega)&=&\chi_{\parallel\parallel\perp}E_{\parallel}(\mathbf{r},\omega)E_{\perp}(\mathbf{r},\omega),\\
\mathbf{P}_{bulk}(\mathbf{r},2\omega)&=&\gamma_{bulk}\mathbf{\nabla}.\left[\mathbf{E}(\mathbf{r},\omega).\mathbf{E}(\mathbf{r},\omega)\right].
\end{eqnarray}
\(\mathbf{E}(\mathbf{r},\omega)\) is the electric field at the fundamental frequency. \(\perp\) and \(\parallel\) denote the normal and tangential components with respect to the particle surface. \(\chi_{\perp\perp\perp}\), \(\chi_{\parallel\parallel\perp}\) and \(\gamma_{bulk}\) are the local surface and nonlocal bulk susceptibilities for a centrosymmetric medium \cite{Wang2009,Bachelier2010}. The SHG being a coherent process, the three contributions interfere: the total SH intensity scattered by the NPs is given by the coherent sum of the electric fields generated by each contribution. It can then be written as follows \cite{Bachelier2010}:
\begin{eqnarray}
I_{SHG}= && G \left|a\mathbf{E}_{surf,\perp}(\mathbf{r},2\omega)+b\mathbf{E}_{surf,\parallel}(\mathbf{r},2\omega) \right. \nonumber \\ 
&&\left. +d\mathbf{E}_{bulk}(\mathbf{r},2\omega)\right|^2
\end{eqnarray}
where \(a\), \(b\) and \(d\) are the Rudnick and Stern parameters linked to \(\chi_{\perp\perp\perp}\), \(\chi_{\parallel\parallel\perp}\) and \(\gamma_{bulk}\), respectively~\cite{Rudnick1971}, and \(G\) is a normalizing constant. \(\mathbf{E}_{surf,\perp}\), \(\mathbf{E}_{surf,\parallel}\) and \(\mathbf{E}_{bulk}\) are harmonic fields associated with the source terms given by Eqs.~(1-3) as obtained from FEM simulations in the far-field region. 

\begin{figure}[tp] 
\center 
\resizebox*{8cm}{!}{\includegraphics{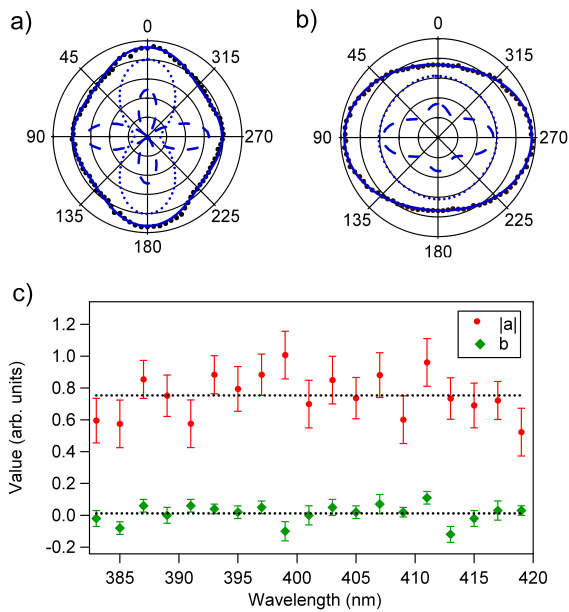}}
\caption{\label{Figure2} (a and b) Normalized SH intensities recorded at a harmonic wavelength of 385~nm and 413~nm, respectively. The experimental data are fitted with Eq.~4 for 140 nm silver spheres (full lines). They are decomposed into the \(\gamma_{bulk}\) (dotted lines) and \(\chi_{\perp\perp\perp}\) (dashed lines) contributions. (c) The norm of the parameter \(a\) (circles) and the value of \(b\) (diamonds) normalized to that of d are reported as functions of the harmonic wavelength. The corresponding fits with constant values are given by the dotted lines.}
\end{figure}
 
As in Ref.~\onlinecite{Bachelier2010}, the intensities recorded for a polarization parallel and perpendicular to the scattering plane were simultaneously fitted by Eq.~4 where the Rudnick and Stern parameters are adjustable variables. The corresponding contributions of the local surface (dashed line) and nonlocal bulk (dotted line) sources are shown in Figs.~2a and~2b for a harmonic polarization into the scattering plane (note that the absolute intensities have been rescaled for clarity). Clearly, the interference pattern between dipolar and octupolar terms significantly depends on the origin of the nonlinearity. This is mainly due to the fact that the retardation effects do not affect the local surface and nonlocal bulk contributions in the same way: the gradient operator in the bulk polarization (Eq.~3) favors the retardation effects at the excitation step. This explains why the octupolar term is weaker in the scattered field (emission step) than for the surface term. As a consequence, the local surface contribution (dashed curves) oscillates more than the corresponding nonlocal bulk contribution (dotted curves). 

The interference pattern also strongly evolves as a function of wavelength close to the octupolar resonance, even if the total intensity shows smooth variations only (compare Figs.~2a and~2b). This wavelength dependence might in principle have two origins: the field enhancements and the dispersion of the Rudnick and Stern parameters. To decide about this origin, the norm of the parameter \(a\) and the value of \(b\) normalized to that of \(d\) are shown as functions of the harmonic wavelength (Fig. 2c). Despite the presence of the quadrupolar and octupolar resonances, both \(|a|\) and \(|b|\) are found to be almost independent of the harmonic wavelength, as expected below the interband threshold for a rather small wavelength excursion. The wavelength dependence of the polarization response is therefore only induced by the resonance effects on the electric fields. Performing the same fitting procedure for smaller (130 nm) or larger (150 nm) nanospheres was not possible owing to the sharp dependence of the resonance wavelengths with particle diameters. This is in agreement with a mean particle size of 140~nm as obtained from the extinction measurements (see Fig.~1a). Though, the phase of the parameter \(a\) smoothly increases by \(\pi/4\) over the wavelength range studied. This results from the fact that we do not take into account the size distribution in the FEM simulations, mainly for computational time reasons. Nevertheless, the fitting procedures lead to \(|a|=0.75\pm0.01\) and \(b=0.01\pm0.01\), demonstrating that the \(\chi_{\parallel\parallel\perp}\) contribution to the total SH intensity is negligible and will therefore be disregarded in the following. In addition, we will considered that \(a\) and \(d\) are constant below the interband threshold, allowing extending the simulations down to 325 nm for the harmonic wavelength, in order to cover entirely the octupolar SPR.

\begin{figure}[tp] 
\center 
\resizebox*{7cm}{!}{\includegraphics{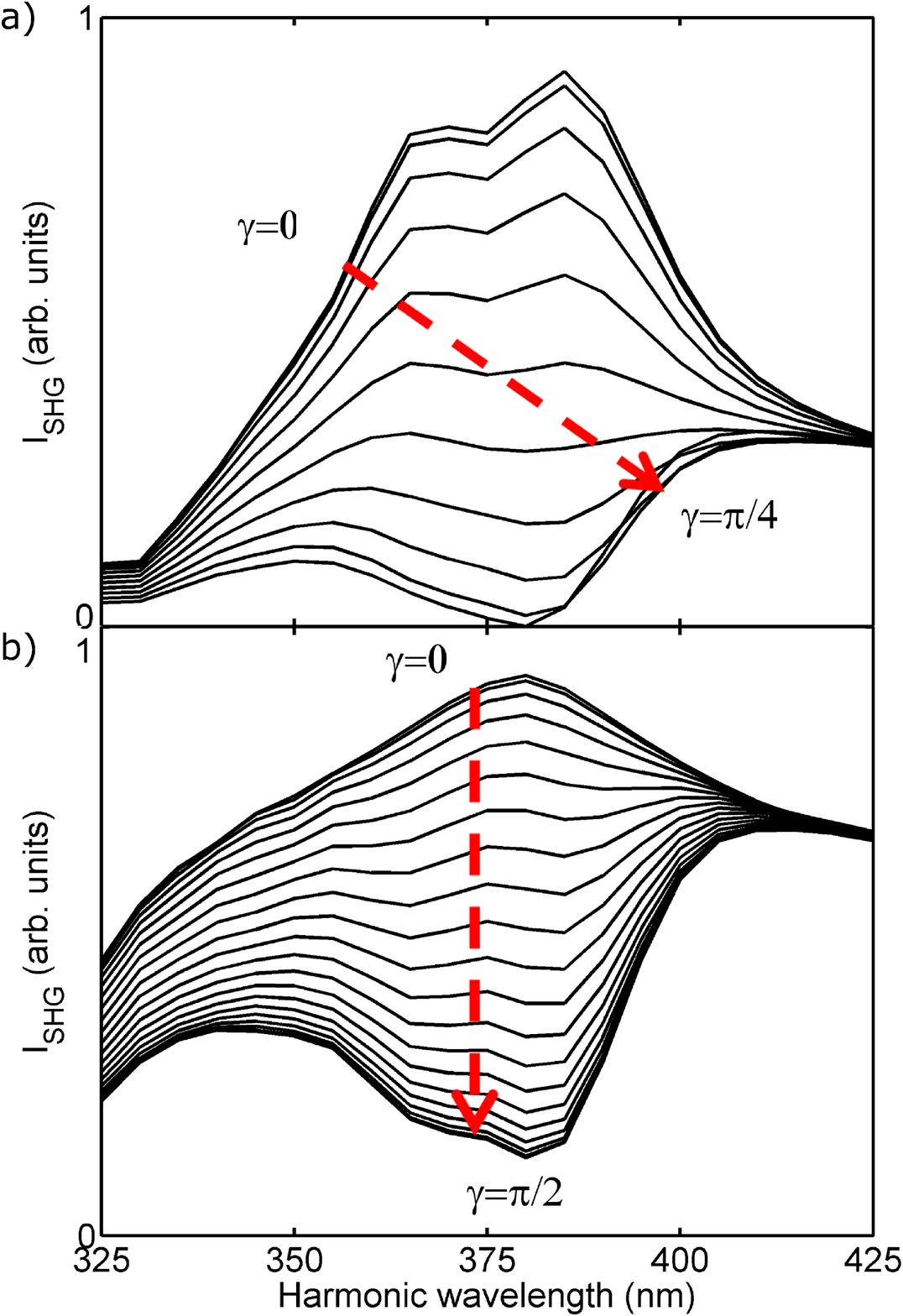}}
\caption{\label{Figure3}Normalized SH intensity for a harmonic wave polarized into the scattering plane as function of the harmonic wavelength and considering the sources \(\chi_{\perp\perp\perp}\) (a) and \(\gamma_{bulk}\) (b) to the nonlinear response. The incident polarization angle is turned from 0 to \(\pi/4\) and to \(\pi/2\) for the curves (a) and (b), respectively.}
\end{figure}

We now have gathered all necessary ingredients to turn to the nonlinear Fano profiles. As shown in Fig.~3, the SH intensity for a polarization into the scattering plane strongly depends on the harmonic wavelength while scanning the octupolar resonance. But more interestingly, the spectral lineshape is extremely sensitive to the incident polarization angle \(\gamma\). In particular, it evolves from a quasi symmetric profile for \(\gamma=0\) to a strongly asymmetric one for \(\gamma=\pi/4\), if one considers the local surface source \(\chi_{\perp\perp\perp}\) only (Fig. 3a). These profiles, revealed for the first time in the SH response of noble metal NPs, are indeed typical for Fano resonances \cite{Fano1961}, which were reported in linear optics of plasmonic NPs only recently \cite{Hao2008,Bachelier2008P}. The main ingredient required for their appearance is the presence of a discrete state coupled to a continuum of states. In the present case, the asymmetric line-shapes arise from the interaction between the broad tail of the dipolar SPR and the relatively narrow octupolar SPR (see Fig.~1). Hence, the retardation effects associated with large particle sizes play a key role here due to the large radiative shift and broadening of the low order SPRs compared to that of higher order SPRs. In that sence, this phenomenon is the nonlinear extension of the Fano profiles observed in the light scattering by finite obstacles reported in Ref.~\onlinecite{Tribelsky2008}. 

The spectral lineshapes shown in Fig.~3a can be fully understood by considering the radiation patterns of the multipolar plasmons: they are driven by the relative weight of the dipolar and octupolar plasmons contributions as a function of the polarization angle \(\gamma\). More precisely, the modes associated with \(m=0\) equally contribute to the scattered field whatever the value of \(\gamma\), whereas those associated with \(m=\pm 2\) exhibit a four-lobe pattern with alternative signs. For the local surface source \(\chi_{\perp\perp\perp}\), the contribution of the \(m=\pm 2\) octupolar mode is large at \(\gamma=0\;[\pi/2]\), leading to a quasi symmetric profile. In contrast, a clear asymmetric profile is observed for \(\gamma=\pi/4\;[\pi/2]\) as the \(m=\pm 2\) octupolar mode cancels out (see Fig. 3a). In this case, this typical Fano lineshape is due to the interactions between the \(m=0\) dipolar and \(m=0\) octupolar modes. As a consequence, the nonlinear Fano profiles can be externally and macroscopically controlled. This singular property has already been demonstrated in linear optics with heterogeneous dimers composed of silver and gold particles \cite{Bachelier2008P}. 

A singular behaviour is observed for the local surface source \(\chi_{\perp\perp\perp}\) at \(\gamma=\pi/4\) (Fig.~3a): the SH intensity completely vanishes for a harmonic wavelength of roughly 380~nm. This comes from the fact that the \(m=0\) dipolar and octupolar modes have exactly the same amplitude but opposite phases, owing to the resonance effects. This complete SHG suppression is a nonlinear version of the plasmon induced transparency revealed in linear optics \cite{Zhang2008,Lukyanchuk2010,Liu2010}, where the destructive interferences between sub- and super-radiant modes silence the nanostructure optical properties. For the nonlocal bulk source \(\gamma_{bulk}\) (Fig.~3b), the contribution of the octupolar modes is weaker, leading to an incomplete SH suppression, even with a polarization angle of \(\gamma=\pi/2\) for which the sign of the electric field is reversed. Owing to the strong sensitivity of nonlinear optics on environment or local morphology modifications, the observed Fano profiles and plasmon-induced SHG transparency are foreseen to have applications in sensing nanodevices, especially by using a nonlinear surrounding medium in order to dramatically increase the production rate of harmonic photons.

In summary, the resonance effects on the optical SHG from silver NPs have been studied by polarization resolved HRS and FEM simulations. We have shown that the interferences between the dipolar and octupolar SPRs, revealed in a specific polarization configuration, lead to nonlinear Fano-profiles that can be externally controlled by the incident polarization. As for linear optics, plasmon-induced transparency is observed in the SHG in noble-metal nanostructures, paving the way to applications in sensing and opening a new branch in nonlinear plasmonics. 

\acknowledgments{S. Huant is acknowledge for a critical reading of the manuscript.}

\end{document}